\documentclass[aps,pra,reprint]{revtex4-1} 

\usepackage[colorlinks,linkcolor=blue,citecolor = blue,urlcolor=blue]{hyperref}
\usepackage{graphicx}
\usepackage{amsmath,amsfonts,amssymb,bm,mathrsfs}
\usepackage{xcolor}
\usepackage{ulem}
\usepackage{gensymb}
\usepackage[percent]{overpic}

\usepackage{braket}

\DeclareMathOperator{\sinc}{sinc}

\begin{document}
\title{Gaussian mode coupling of spectrally broadband photons from bulk spontaneous parametric down-conversion: A spatial-spectral mode analysis of fiber coupling}

\author{Carlos Sevilla-Guti\'errez}\email{carlos.sevilla@iof.fraunhofer.de}
\affiliation{Fraunhofer Institute for Applied Optics and Precision Engineering, 
07745 Jena, Germany}
\affiliation{Friedrich Schiller University Jena, Institute of Applied Physics, Abbe Center of Photonics, 
07745 Jena, Germany}
\author{Varun Raj Kaipalath}
\affiliation{Fraunhofer Institute for Applied Optics and Precision Engineering, 
07745 Jena, Germany}
\affiliation{Friedrich Schiller University Jena, Institute of Applied Physics, Abbe Center of Photonics, 
07745 Jena, Germany}
\author{Fabian Steinlechner}\email{fabian.steinlechner@uni-jena.de}
\affiliation{Fraunhofer Institute for Applied Optics and Precision Engineering, 
07745 Jena, Germany}
\affiliation{Friedrich Schiller University Jena, Institute of Applied Physics, Abbe Center of Photonics, 
07745 Jena, Germany}
\date{\today}

\begin{abstract} 
Photon sources based on spontaneous parametric down-conversion (SPDC) are central to experimental quantum optics and quantum technologies. Their performance is commonly quantified by three  metrics: pair-collection probability, heralding efficiency, and spectral purity. In bulk-crystal SPDC, these metrics are known to be mutually constrained, yet the physical origin of the resulting trade-offs is often obscured. We show that these trade-offs originate from the frequency-dependent population of discrete spatial modes in the SPDC emission. By performing a Laguerre-Gauss mode decomposition at each frequency component, we show how spectral-spatial non-separability impacts collection probability, heralding efficiency, and purity. We apply this framework to two widely used quasi-phase-matching configurations: collinear degenerate type-0 and type-II SPDC in periodically poled bulk crystals, and quantify how different phase-matching functions shape the spectral-spatial mode structure. In particular, for type-II SPDC we compare standard periodically poled and aperiodically poled Gaussian phase matching. We experimentally validate some of our theoretical results using spatial- and spectral-projection measurements. This spectral-spatial mode analysis provides a quantitative and predictive framework for understanding and engineering bulk-crystal photon sources, enabling systematic multi-parameter optimization beyond qualitative design guidelines.

\end{abstract}

\pacs{} \maketitle

\section{Introduction}

Spontaneous parametric down-conversion (SPDC) is a central source of nonclassical light, producing squeezed vacuum states \cite{Wu:1985_Squeezing_SPDC} and enabling heralded single-photon generation \cite{helraded_single_photon_2016} and entanglement across multiple degrees of freedom \cite{Roux:2020_entanglement}. Many applications require SPDC sources that are both bright and efficiently coupled into a single spatial mode defined by an optical fiber, as in loophole-free Bell tests \cite{Steinlechner:2015_loophole}, boson sampling \cite{Pan:2021_Pan,Zhang:2018_Scattershot}, and quantum sensing \cite{uncoditional:2024_Jian}. While bulk nonlinear crystals avoid the coupling and propagation losses of current waveguide sources \cite{Kashiwasaki:2021_WG_loss}, they intrinsically generate spatially multimode emission, in conflict with the single-mode requirements of most experiments.

\begin{figure}
    \centering
    \includegraphics[width=0.8\linewidth]{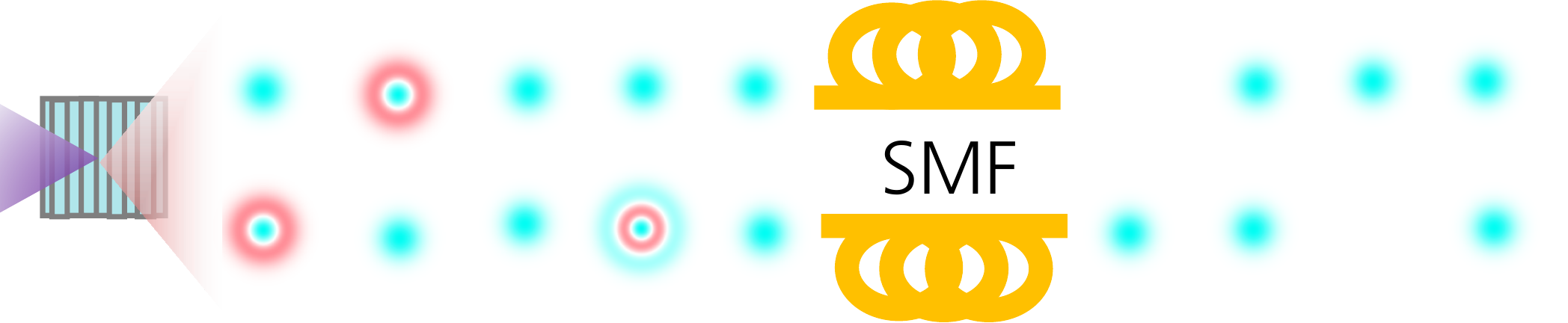}
    \caption{Conceptual sketch of heralding efficiency for SPDC photons coupled into single-mode fiber. A Gaussian pump beam drives SPDC in a nonlinear crystal, generating photon pairs in different spatial mode combinations. Blue spots indicate events in which at least one photon occupies the fundamental Gaussian mode and is therefore coupled into the fiber. Events where both photons populate higher-order spatial modes are not shown, as they do not contribute to the detected single- or coincidence detections.}
    \label{fig:concept}
\end{figure}

Extensive work has sought to analyze the beam parameters of pump- and collection modes that maximize the \textit{pair-collection probability}, i.e. the probability that both photons are emitted into Gaussian modes\,\cite{Minozzi:13,Kurtsiefer:01,Bovino:03,Dragan:04,Andrews:04,Ljunggren:05,Fedrizzi:07,Ling:08,Bennink:10,Palacios:11,Guerreiro:13}.  This metric can be normalized either with respect to the pump photons incident on the crystal, yielding the power-normalized brightness (pairs/mW), or as the conditional collection probability, i.e., the fraction of photons collected given that a pair was generated in the SPDC process. Although the latter is experimentally difficult to access, it more directly reflects the spatial characteristics of the emission. In the following, we adopt the latter definition. An equally--if not experimentally even more--important figure of merit is the \textit{heralding efficiency} (also known as the \textit{Klyshko efficiency}), defined as the conditional probability of detecting a photon given the detection of its partner. While the pair collection probability quantifies how often both photons are simultaneously emitted into Gaussian modes, the heralding efficiency measures how reliably the detection of one photon heralds the presence of its partner in the same mode. Figure~\ref{fig:concept} illustrates the distinction: pair collection counts the subset of events where both photons occupy the Gaussian mode (simultaneous SMF coupling), while heralding also includes events where only one photon is Gaussian; detection of that photon heralds its partner, which may be rejected by the fiber if it occupies a higher-order mode.

Previous studies \cite{Kurtsiefer:01,Bovino:03,Dragan:04,Andrews:04,Ljunggren:05,Fedrizzi:07,Ling:08,Bennink:10,Palacios:11,Guerreiro:13,SteinlechnerPhD:2015} have shown that these two figures of merit cannot be optimized simultaneously: tight focusing enhances pair collection at the expense of heralding, while loose focusing has the opposite effect. Bennink  \cite{Bennink:10} provided a detailed analysis of this trade-off for non-degenerate quasi-phase-matching configurations, which were later experimentally corroborated  \cite{Dixon_2014,SteinlechnerPhD:2015}. Building on this work, heralding efficiencies exceeding $0.8$ have been demonstrated, including overall system losses \cite{Ramelow:13_heralding,Slussarenko:2017_metrology,You:2021_metrology, Steinlechner:2015_loophole}. However, these analyses treated heralding and pair-collection efficiencies as abstract optimization targets, offering limited insight into the underlying physical mechanism. Related insights were obtained by Osorio \textit{et al.}\,\cite{Osorio2008} who studied spatial purity under spatial and spectral intensity filtering. They showed that tighter transverse momentum filtering increases the spatial purity of the SPDC state, which allows for a Schmidt-mode decomposition of the biphoton field  \cite{Eberly:2004,Miatto_schmidt:2012}. In this limit, perfect heralding efficiency is, in principle, attainable. Guerreiro \textit{et al.}\,\cite{Guerreiro:13} linked the spectral dependence of the emission angles often called X-entanglement\,\cite{Gatti_Xentanglement} to fiber coupling by matching the angular divergence to that of a Gaussian collection mode. However, these approaches does not provide a quantitative description of how the frequency-dependent excitation of spatial modes mutually constrains pair-collection, heralding, and spectral purity.

Here we develop a quantitative framework that resolves the frequency dependence of the spatial mode structure of SPDC \cite{Sevilla:2024_SpectralLG,Baghi:2022_Generalized}. A frequency-resolved decomposition of the biphoton state into Laguerre-Gauss (LG) modes, with emphasis on the radial modes, reveals how spectral-spatial non-separability impacts pair-collection probability and heralding efficiency when integrated over finite detection bandwidths. We show that related trade-offs arise from the spectral distinguishability of different spatial mode combinations, which is strongly controlled by the focusing conditions.

 Our work extends the seminal analysis of Bennink \cite{Bennink:10} in two important directions. First, we analyze pair-collection and heralding efficiencies for additional quasi-phase-matching configurations now widely employed in experiments, in particular degenerate type-0 SPDC and type-II SPDC in aperiodically poled crystals. Second, rather than treating these efficiencies as abstract optimization targets, we identify their physical origin and mutual constraints induced by spectral-spatial coupling.

In addition, we show how this framework guides the optimization of competing source performance parameters. For the type-II quasi-phase-matching configuration, we consider aperiodic poling in domain-engineered nonlinear crystals \cite{Graffitti2017}, now a standard approach for generating spectrally pure photons, and show that -- with regards to spatial mode profile -- behavior analogous to periodic poling is recovered after appropriate scaling of the focusing parameters. Finally, we quantify how focusing impacts the spectral purity and identify a threshold value of the focusing parameter below which the purity remains close to unity.

The remainder of the paper is organized as follows. In Section~\ref{sec:theory}, we review the frequency-dependent Laguerre-Gauss mode description of SPDC and the definitions of spectral purity, pair-collection probability, and heralding efficiency. Section~\ref{sec:Methods} details the numerical parameters used in our simulations and further considerations. In Section~\ref{sec:results}, we highlight their distinct trade-offs for type-II and type-0 quasi-phase-matching configurations and relevant optimization strategies. Furthermore, we show experimental validation of some of our theoretical results. Finally, we summarize our conclusions in Section~\ref{sec:discussion}.

\section{Theory}\label{sec:theory}

The SPDC biphoton state can be expanded in the LG basis as\,\cite{Baghi:2022_Generalized}

\begin{align}
 \label{eq:LG_basis_freq}
\ket{\Psi}=  N\iint d\Omega_\mathrm{s}d\Omega_\mathrm{i}
 \sum_{p_\mathrm{s},p_\mathrm{i}=0}^{\infty} \sum^{\infty}_{\ell_\mathrm{s},\ell_\mathrm{i}=-\infty}C_{p_\mathrm{s},p_\mathrm{i}}^{\ell_\mathrm{s},\ell_\mathrm{i}}(\Omega_\mathrm{s}, \Omega_\mathrm{i}) &\\ \nonumber \times \hat{A}^\dagger_{p_\mathrm{s},\ell_\mathrm{s}}(\Omega_\mathrm{s})\hat{A}^\dagger_{p_\mathrm{i},\ell_\mathrm{i}}(\Omega_\mathrm{i})\ket{Vac} 
\end{align}

where $\Omega_\mathrm{j}$ denotes the frequency detuning from the central frequencies $\omega_\mathrm{j,0}$, i.e.\ $\omega_\mathrm{j}=\omega_\mathrm{j,0}\pm\Omega_\mathrm{j}$ ($\mathrm{j}=\mathrm{s},\mathrm{i}$), and $N$ is a normalization constant. The operators $\hat{A}^\dagger_{p_\mathrm{j},\ell_\mathrm{j}}$ create photons in LG modes characterized by radial index $p_\mathrm{j}$ and azimuthal index $\ell_\mathrm{j}$. The spectral amplitude 

\begin{align}\label{coe1}
    C^{\ell_\mathrm{s},\ell_\mathrm{i}}_{p_\mathrm{s},p_\mathrm{i}}(\Omega_\mathrm{s}, \Omega_\mathrm{i}) 
    =    \iint  d\bm{q}_\mathrm{s} \: d\bm{q}_\mathrm{i}\int^{L/2}_{-L/2}dz
    \,\chi^{(2)}(z) \exp{(-i\Delta k z)}\nonumber&\\
    \times F(\Omega_\mathrm{p})\mathrm{LG}_{0}^{0}(\bm{q}_\mathrm{p})[\mathrm{LG}_{p_\mathrm{s}}^{\ell_\mathrm{s}}(\bm{q}_\mathrm{s})]^*         [\mathrm{LG}_{p_\mathrm{i}}^{\ell_\mathrm{i}}(\bm{q}_\mathrm{i})]^*,
\end{align}

encodes the joint spectral amplitude (JSA) for signal and idler photons occupying specific spatial modes. Here, we have considered a spatial Gaussian mode for the pump field $\mathrm{LG}_{0}^{0}(\bm{q}_\mathrm{p})$ with an arbitrary spectral amplitude $F(\Omega_\mathrm{p})$. Furthermore, $\bm{q}_\mathrm{j}$ denote transverse momentum vectors, $\chi^{(2)}(z)$ is the (possibly $z$-dependent) second order nonlinearity, and $\Delta k = k^z_\mathrm{p}- k^z_\mathrm{s} - k^z_\mathrm{i}-\frac{2\pi}{\Lambda}$ is the longitudinal phase mismatch, with $\Lambda$ being the poling period. The latter depends on the transverse momentum and frequency of the photons and under a Taylor expansion can be written as\,\cite{Baghi:2022_Generalized}

\begin{align}\label{eq:Long_Ph_mis_gen}
    \Delta k= \Delta k_0 + (\frac{1}{u_{\mathrm{p}}}-\frac{1}{u_{\mathrm{s}}})\Omega_\mathrm{s}+(\frac{1}{u_{\mathrm{p}}}-\frac{1}{u_{\mathrm{i}}})\Omega_\mathrm{i}+\frac{G_\mathrm{s}}{2}\Omega_\mathrm{s}^2&\\+\frac{G_\mathrm{i}}{2}\Omega_\mathrm{i}^2+\frac{|\bm{q}_\mathrm{s}+\bm{q}_\mathrm{i}|^2}{2k_{\mathrm{p},0}}\nonumber-\frac{|\bm{q_\mathrm{s}}|^2}{2k_{\mathrm{s},0}}-\frac{|\bm{q_\mathrm{i}}|^2}{2k_{\mathrm{i},0}}.
\end{align}

Here, $u_\mathrm{j}$ and $G_\mathrm{j}$ denote the group velocity and group-velocity dispersion, respectively, and $k_{\mathrm{j},0}$ are the wavenumbers evaluated at the central frequencies $\omega_{\mathrm{j},0}$. The contribution $\Delta k_0=k_{\mathrm{p},0}- k_{\mathrm{s},0} - k_{\mathrm{i},0}-2\pi/\Lambda$ is the phase-mismatch at the central frequencies. It is chosen to vanish for an initially phase-matched configuration, but can be tuned, for example, through the temperature dependence of the refractive indices and the thermal expansion of the poling period\,\cite{Nina_cryo_2023,Lerch:13_tuningT0,Song:25_temp_pump}. We write this contribution as $\Delta k_0(T)=2\phi(T)/L$, where $\phi(T)$ is a dimensionless parameter. In the following, the explicit temperature dependence is omitted for brevity. This parameter plays an important role when optimizing Gaussian-mode overlap, since focusing introduces an additional phase contribution \cite{Boyd:68,Bennink:10}.

The spectral purity $\mathcal{P}$ can be determined by performing Schmidt decomposition of the JSA to extract the Schmidt coefficients $\{\sqrt{\lambda_k}\}$. It is then given by\,\cite{Ansari:18}
\begin{equation}
    \mathcal{P}=\sum_k\lambda_k^2
\end{equation}

In this work we restrict our attention to cases where at least one photon occupies the fundamental Gaussian mode. Although the fundamental mode of a SMF is the linearly polarized mode $LP_{0,1}$, it is very well approximated by a Gaussian mode\,\cite{Bruning:15}. Moreover, due to orbital angular momentum conservation in SPDC ($\ell_\mathrm{p}=\ell_\mathrm{s}+\ell_\mathrm{i}$) \cite{Mair:01}, we can neglect all terms with $|\ell_\mathrm{j}|>0$ and henceforth omit the azimuthal index.

The dependence on pump, signal, and idler waists enters through the LG mode functions (see Appendix). Following Bennink \cite{Bennink:10}, we characterize focusing using the dimensionless parameter $\xi=L\lambda/(2\pi n w^2)$, where $w$ is the beam waist, $L$ is the crystal length, $\lambda$ is the wavelength and $n$ is the refractive index. 

The pair-collection probability integrated over the detection bandwidth is given by:

\begin{equation}
    S^{(2)}=N^2\iint d\Omega_\mathrm{s}d\Omega_\mathrm{i} P_{0,0}(\Omega_\mathrm{s},\Omega_\mathrm{i})\, ,
\end{equation}

where $P_{p_\mathrm{s},p_\mathrm{i}}(\Omega_\mathrm{s},\Omega_\mathrm{i})=|C_{p_\mathrm{s},p_\mathrm{i}}(\Omega_\mathrm{s},\Omega_\mathrm{i})|^2$ denotes the spectral density for mode combinations with radial indices $p_\mathrm{j} (\mathrm{j}=\mathrm{s},\mathrm{i})$. From now on, for brevity, we omit the spectral dependence of $P_{p_\mathrm{s},p_\mathrm{i}}$.The corresponding single-photon detection probabilities are

\begin{align}
        S^{(1)}_\mathrm{s} =N^2\iint d\Omega_\mathrm{s}d\Omega_\mathrm{i}\sum_{p_\mathrm{i}} P_{0,p_\mathrm{i}},\,&\\S^{(1)}_\mathrm{i} =N^2\iint d\Omega_\mathrm{s}d\Omega_\mathrm{i} \sum_{p_\mathrm{s}}P_{p_\mathrm{s},0}\nonumber\,.
\end{align}

In general, $P_{x,0}\neq P_{0,x}$ for type-II and strongly non-degenerate SPDC, reflecting asymmetries arising from different refractive indices and emission angles.

The heralding efficiencies are defined as

\begin{equation}\label{eq:decomposition_gauss}
    H_\mathrm{s} =  \frac{S^{(2)}}{S^{(1)}_{\mathrm{i}}} , \,\,\, H_\mathrm{i} =  \frac{S^{(2)}}{S^{(1)}_{\mathrm{s}}} \,. 
\end{equation}

and we further introduce the symmetric heralding efficiency $H=\sqrt{H_\mathrm{s}H_\mathrm{i}}$, which will be used throughout the remainder of this work. In experiments, heralding efficiencies are typically inferred from coincidence-to-single detection ratios and include additional system losses.

\section{Methods}\label{sec:Methods}

Throughout this work, we use the standard nomenclature for SPDC phase matching: type-0 refers to co-polarized pump, signal, and idler fields; type-I to co-polarized signal and idler fields orthogonal to the pump; and type-II to signal and idler photons of orthogonal polarization. We treat type-II and degenerate type-0 quasi-phase-matching configurations separately, as their phase-matching properties lead to qualitatively different efficiency trade-offs. These differences can be understood from the frequency dependence of the longitudinal phase mismatch that can be seen in Eq.\,\ref{eq:Long_Ph_mis_gen}. 

For type-II SPDC (and strongly non--degenerate type-0/I), the group-velocity mismatch dominates, such that the linear term in $\Omega$ governs the phase matching. In contrast, for degenerate type-0 SPDC the group velocities are equal, and the quadratic dispersion term determines the spectral structure. As a consequence, degenerate type-0 configurations are sensitive to residual phase mismatch $\phi$, whereas in the type-II regime the pair-collection probability $S^{(2)}$ and heralding efficiency $H$ are independent of $\phi$.  

These differences are also reflected in the scaling of the brightness with crystal length $L$. Here we mean the photon-pair generation rate in the Gaussian mode integrated over the full spectral bandwidth. In the linear-dispersion-dominated regime, $S^{(2)}$ is independent of $L$, whereas for purely quadratic dispersion it scales as $S^{(2)} \propto \sqrt{L}$ in bulk crystals for fixed focusing parameters. Note that the scaling is modified in the presence of spectral filtering or waveguide confinement. The derivation of these scaling laws, together with a summary of the relevant scaling behaviors for different parameter regimes, is provided in Appendix~\ref{app_scaling}.

For simplicity, we consider the case of a monochromatic pump ($F(\Omega_\mathrm{s},\Omega_\mathrm{i})=\delta(\Omega_\mathrm{s}+\Omega_\mathrm{i})$) for the calculation of the heralding efficiency and pair-collection probability. With this assumption, it is easy to see that $\Omega_\mathrm{s}=-\Omega_\mathrm{i}=\Omega$ and Eq.\,\ref{eq:Long_Ph_mis_gen} can be written as:

\begin{align}\label{eq:Long_Ph_mis_CW}
    \Delta k_{CW}=\frac{2\phi}{L} +D\Omega+G\Omega^2+\frac{|\bm{q}_\mathrm{s}+\bm{q}_\mathrm{i}|^2}{2k_{\mathrm{p},0}}-\frac{|\bm{q_\mathrm{s}}|^2}{2k_{\mathrm{s},0}}-\frac{|\bm{q_\mathrm{i}}|^2}{2k_{\mathrm{i},0}}.
\end{align}

\noindent

where $D=1/u_\mathrm{s}-1/u_\mathrm{i}$ and $G=(G_\mathrm{s}+G_\mathrm{i})/2$. Varying the pump frequency introduces an effective phase mismatch; thus, for the case of type-II (or highly non-degenerate type-0 and type-I), our results are applicable for the case of a pulsed pump. Note that this is not the case for degenerate type-0 as the heralding efficiency and pair-collection probability dependent on the phase mismatch.

However, the spectral purity $\mathcal{P}$ is highly dependent on the spectral properties of the pump beam and the JSA. For this, we will consider the case of symmetric group velocity matching (SGVM), which is met when $2/u_\mathrm{p}=1/u_\mathrm{s}+1/u_\mathrm{i}$\,\cite{Ansari:18}, achievable in potassium titanyl phosphate (KTP) crystals at the telecom wavelength. Inserting it in Eq.\,\ref{eq:Long_Ph_mis_gen} yields

\begin{align}\label{eq:Long_Ph_mis_SGVM}
    \Delta k_{SGVD}= \frac{2\phi}{L}  + \frac{D}{2}(\Omega_\mathrm{i}-\Omega_\mathrm{s})+\frac{G_\mathrm{s}}{2}\Omega_\mathrm{s}^2+\frac{G_\mathrm{i}}{2}\Omega_\mathrm{i}^2\nonumber&\\+\frac{|\bm{q}_\mathrm{s}+\bm{q}_\mathrm{i}|^2}{2k_{\mathrm{p},0}}-\frac{|\bm{q_\mathrm{s}}|^2}{2k_{\mathrm{s},0}}-\frac{|\bm{q_\mathrm{i}}|^2}{2k_{\mathrm{i},0}}.
\end{align}

Note that, up to the first-order dispersion, Eq.\,\ref{eq:Long_Ph_mis_SGVM} exhibits the same spectral dependence as Eq.\,\ref{eq:Long_Ph_mis_CW}. Therefore, we can obtain the corresponding expression by rotating Eq.\,\ref{eq:Long_Ph_mis_CW} by $45\deg$ in the $(\Omega_\mathrm{s}, \Omega_\mathrm{i})$-plane to calculate $\mathcal{P}$. 

Rather than reporting the absolute pair-collection probability per pump photon, we analyze how the brightness varies with focusing parameters and phase mismatch relative to the optimal configuration. We therefore introduce the \textit{relative brightness}

\begin{equation}
B=S^{(2)}/\mathrm{max}_{\xi_\mathrm{p},\xi_\mathrm{s},\xi_\mathrm{i},\Phi}S^{(2)}\,.
\end{equation}

For simplicity, we set $\xi_\mathrm{s}=\xi_\mathrm{i}$, since in the degenerate case the momentum correlations are symmetric with respect to the signal and idler transverse momenta, leading to modes with similar angular spread. This approximation remains well motivated for degenerate type-II SPDC, where the birefringence only weakly perturbs this symmetry. However, highly nondegenerate configuration breaks this symmetry and need to be considered more carefully and, therefore, is outside of the scope of our work. We further elaborate on this point in Appendix\,\ref{sec:appen_nondeg}. We restrict the focusing parameters to $\xi\leq10$, covering the range relevant for most experimental implementations. As our primary interest lies in high-heralding-efficiency operation, we focus on the regime $H>0.7$, which allows truncation of the radial mode expansion at $p_\mathrm{j}=4$. Higher-order terms contribute less than $0.02$ to $H$ at this threshold and decrease rapidly for increasing heralding efficiency.

All results are obtained numerically. Details of the parameters, ranges, and step sizes used in the simulations including $L$, $\lambda_\mathrm{p,s,i}$, $\xi$, $\phi$, and material properties--are provided in the Appendix. To ensure generality, results are expressed in terms of the dimensionless variables $\xi$, $D\Omega L$, and $\sqrt{G L}\,\Omega$, allowing direct application to different experimental configurations.

With these definitions in place, we first analyze the dependence of the relative brightness and heralding efficiency on the focusing parameters $(\xi_\mathrm{p},\xi_\mathrm{s},\xi_\mathrm{i})$. We then examine the spectral dependence of the coefficients $C_{0,p_\mathrm{s}}(\Omega)$, which provides direct insight into the physical origin of the efficiency trade-offs.

\section{Results}\label{sec:results}
\subsection{Collinear Type-II, periodic poling}\label{subsec:type-II}

\begin{figure}[t]
    \centering
    \begin{overpic}[width=1.0\linewidth]{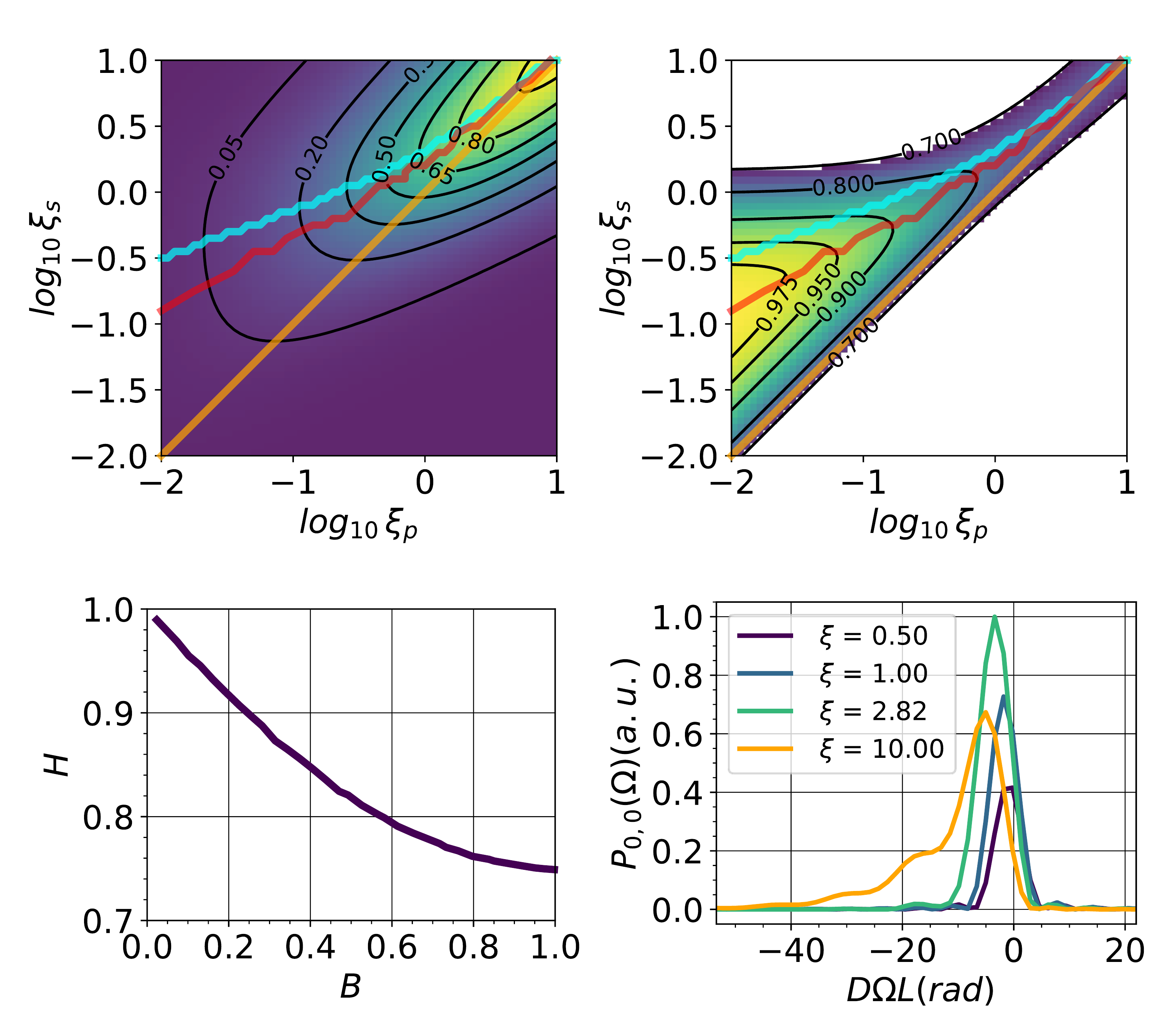}

   \end{overpic}
    
   \caption{(Upper left) Relative brightness $B$ and (upper right) heralding efficiency $H$ as a function of the focusing parameters for pump $\xi_\mathrm{p}$ and SPDC photons $\xi_\mathrm{s}$. Colored lines represent different optimization strategies. The orange/cyan lines are obtained when optimizing $B$ by fixing $\xi_\mathrm{s}$/$\xi_\mathrm{p}$ and optimizing over $\xi_\mathrm{p}$/$\xi_\mathrm{s}$. Finding different trade-off with respect to $H$. The red line depicts the optimization of $H$ by setting a given value of $B$ and optimizing over $\xi_\mathrm{s}$ and $\xi_\mathrm{p}$. (Bottom left) Results of the last optimization, showing the trade-off between $B$ and $H$. Bottom right) Spectral density $P_{0,0}(\Omega)$ for different focusing conditions $\xi_\mathrm{p}=\xi_\mathrm{s}=\xi$. }
    \label{fig:B_H_BvsH_focus_shift}
\end{figure}

Figure~\ref{fig:B_H_BvsH_focus_shift} shows the relative brightness $B$ and heralding efficiency $H$ as functions of the pump and collection focusing parameters $\xi_\mathrm{p}$ and $\xi_\mathrm{s}$. Since $B$ increases asymptotically with tighter focusing, it is normalized to its maximum value within the explored parameter range. The maxima of $B$ and $H$ occur at opposite extremes of the focusing landscape, directly revealing the trade-off between brightness and heralding efficiency.

Three optimization strategies are indicated in Fig.~\ref{fig:B_H_BvsH_focus_shift}. Maximizing $B$ at fixed collection optics yields the well-known condition $\xi_\mathrm{p}=\xi_\mathrm{s}$ (orange curve), while maximizing $B$ at fixed pump focusing leads to the cyan curve. Although both strategies achieve the same relative brightness, the latter provides systematically higher heralding efficiency. Directly optimizing $H$ over  $\xi_p$ and $\xi_s$ at fixed target values of $B$ yields the red curve, which achieves the highest heralding efficiency with only a modest reduction in brightness, making it  the most attractive of the three. The resulting trade-off is summarized in Fig.~\ref{fig:B_H_BvsH_focus_shift} (bottom left), where achieving $H>0.97$ requires a reduction in relative brightness by approximately one order of magnitude, consistent with previous theoretical and experimental results. The fitted functions for $H$, $\xi_\mathrm{p}$ and $\xi_\mathrm{s}$ as functions of $B$ for this optimization strategy are provided in Appendix\,\ref{Appen-II}. 

We now examine how the spectral density $P_{0,0}$ depends on the focusing parameters, which is shown in Fig.~\ref{fig:B_H_BvsH_focus_shift} (bottom right). Increasing the focusing broadens the spectrum and reduces its peak amplitude, while further enhancing the integrated brightness. Notably, the peak spectral density occurs at $\xi_\mathrm{s}=\xi_\mathrm{i}=2.82$, consistent with the Boyd-Kleinman focusing condition for second-harmonic generation \cite{Boyd:68}. In addition to spectral broadening, the spectral density deviates from the $sinc$ shape observed for a plane-wave interaction. Interestingly, when considering the case of SGVM, it can be used purposefully to increase the spectral purity as a result of the reduction of the sinc-function sidelobes, as noted by Bennink\,\cite{Bennink:10}. For stronger focusing the spectrum becomes more asymmetric and the spectral purity decreases. A technical aspect for experimentalists to take into consideration is that the center wavelength may shift considerably with focusing. This must be taken into account by temperature tuning the temperature of the NLC (i.e. tuning $\phi$).

\begin{figure}[t]
    \centering
    \includegraphics[width=1.0\linewidth]{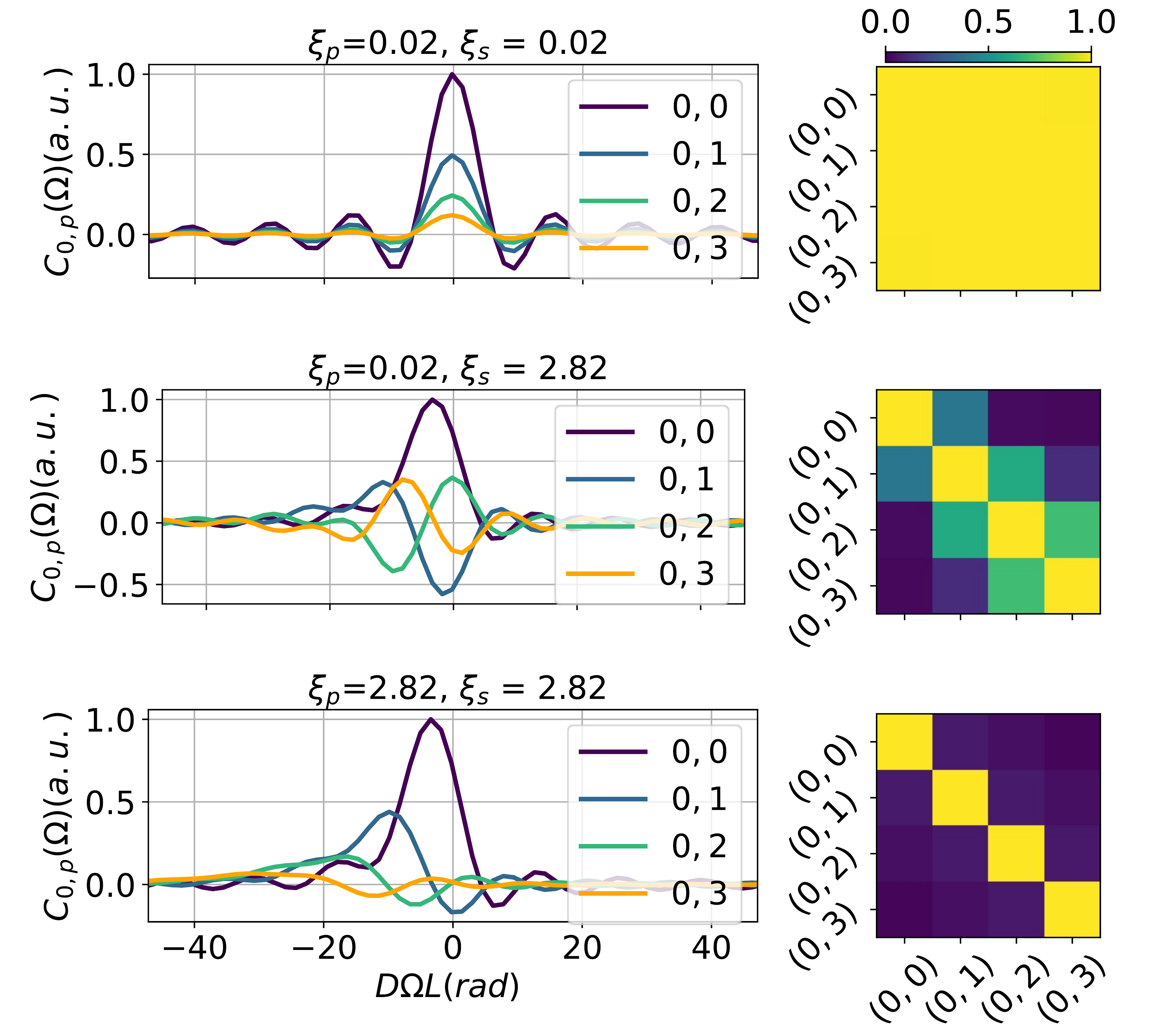}
    \caption{The first column show the spectral amplitudes, $C_{0,p}(\Omega)$, for different joint modes depicting how spectrally distinguishable the different mode combination and for different focusing parameters used. The second column quantifies the spectral similarity by calculating the overlap of the normalized spectral amplitudes  $C'_{0,x}=C_{0,x}/\sqrt{\int d\Omega P_{0,x}}$. The first row shows the case of loose focusing ($\xi_\mathrm{p}=\xi_\mathrm{s}=0.02$) where the spectra overlap is near to unity. The second row a tighter focusing was selected for the SPDC photons ($\xi_\mathrm{p}=2.8$, leading to greater spectral distinguishability. The last column shows how the case of $\xi_\mathrm{p}=\xi_\mathrm{s}=2.8$, where the effect is stronger almost reaching to spectral orthogonality}
    \label{fig:spectrum_Cp_Overlap}
\end{figure}

Now, we look at the spectral dependence of the joint spatial modes that play a role in the heralding efficiency. Figure\,\ref{fig:spectrum_Cp_Overlap} shows the spectral amplitudes, $C_{0,p}$, for $p = 0, 1, 2, 3$, for three different configurations of $\xi_\mathrm{p}$ and $\xi_\mathrm{s}$. On the right-hand side, we present the plot of the overlap between the normalized spectral amplitudes  $C'_{0,x}=C_{0,x}/\sqrt{\int d\Omega P_{0,x}}$, to quantify the similarity between the spectral amplitudes. Figure~\ref{fig:spectrum_Cp_Overlap} reveals how focusing controls the spectral distinguishability of joint spatial modes. For loose focusing of pump and collection, all modes exhibit nearly identical sinc-shaped spectra, leading to near-unity overlap and approximate spectral-spatial separability, $C_{p_\mathrm{i},p_\mathrm{s}}(\Omega)\approx g_{p_\mathrm{i},p_\mathrm{s}}f(\Omega)$. In this regime, $g_{p_\mathrm{i},p_\mathrm{s}}$ can be diagonalized through a Schmidt decomposition, yielding the spatial Schmidt basis for signal and idler photons \cite{Eberly:2004,Miatto_schmidt:2012}. In this basis, unit heralding efficiency is attainable in principle.

With tighter focusing, the spectra of different spatial modes become increasingly distinct, and their overlap decays rapidly, approaching near-orthogonality when $\xi_\mathrm{p}=\xi_\mathrm{s}$. This spectral-spatial entanglement renders the spatial degree of freedom partially mixed and fundamentally limits heralding efficiency achievable by spatial projection alone.

For the same reason, monochromatic (or sufficiently narrowband) filtering yields near-unity heralding efficiency at any $\xi_p$ \cite{SteinlechnerPhD:2015}. However, such narrowband spectral filtering is typically not experimentally viable for type-II SPDC configuration, where the phase-matching bandwidth tends to be on the (sub)nanometer scale (Fig. \ref{fig:experiment_results}). Moreover, in the pulsed-pump regime, spectral filtering can reduce the heralding efficiency, as studied in detail by Meyer et.al\,\cite{Meyer-Scott:2017_helrading_pulsed}.

Together, these results demonstrate that the brightness-heralding trade-off in type-II SPDC originates from frequency-dependent spectral distinguishability between spatial modes, rather than from spatial mode structure alone.

\subsubsection{Experiment}\label{app:experiment}
\begin{figure}[t]
    \centering
    \includegraphics[width=1\linewidth]{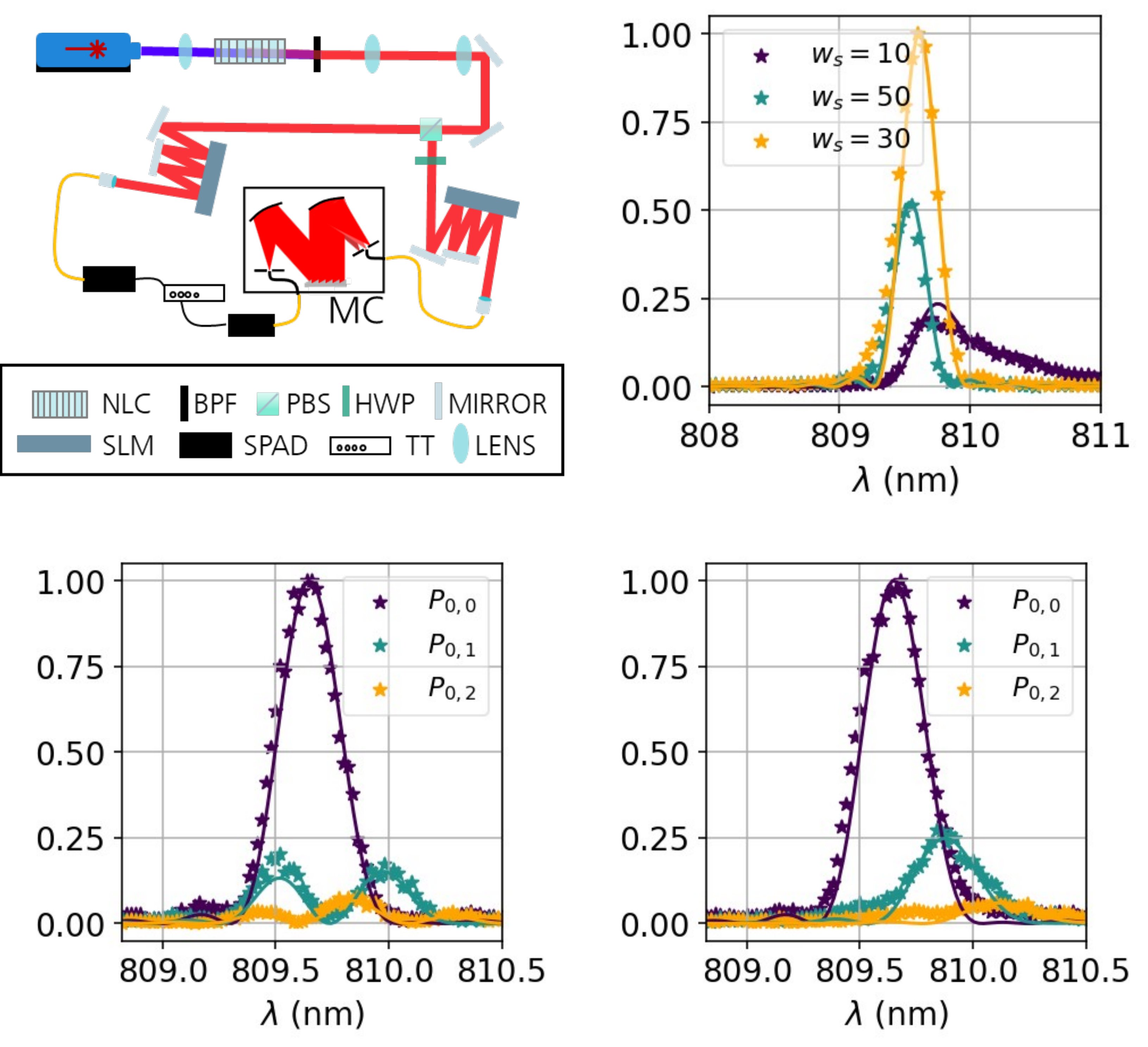}
    \caption{(Upper left) Experimental setup for the spatial--spectral analysis of the SPDC state, using spatial and spectral mode projections via multi-plane light conversion and a monochromator, respectively. MC: Monochromator; NLC: nonlinear crystal; SLM: spatial light modulator; BPF: band-pass filter; SPAD: single-photon avalanche diode; PBS: polarizing beam splitter; HWP: half-wave plate; TT: time tagger. (Upper right) Spectra of joint Gaussian-mode coupling ($p_{\mathrm{s}}=p_{\mathrm{i}}=0$) for different collection waists and fixed pump waist $w_\mathrm{p}=20\,\mu\mathrm{m}$, showing spectral shift and broadening under tighter focusing. (Bottom left) and (bottom right) Spectra of different $(0,p_\mathrm{i})$ modes for collection waists $w_\mathrm{s}=30\,\mu\mathrm{m}$ and $20\,\mu\mathrm{m}$, respectively.}
    \label{fig:experiment_results}
\end{figure}

Here, we test several of our predictions experimentally. Figure\,\ref{fig:experiment_results}\,(upper left) shows the experimental setup used. A 404.8\,nm pump laser is focused at the center of a 20\,mm-long PPKTP nonlinear crystal (NLC). After the SPDC process, the pump is removed with a long-pass filter (LPF), and the central plane of the NLC is imaged onto the spatial-mode detection system using a 4-f system with 10x magnification. The signal and idler photons are separated by a polarizing beam splitter (PBS) and routed to different detection arms. To detect different spatial modes, we employ the multiplane light conversion (MPLC) technique consisting of spatial light modulators (SLMs), three phase planes, and single-mode-fiber coupling\,\cite{Sevilla:2024_SpectralLG}. The MPLC was previously characterized, showing a nearly mode-independent efficiency ($\approx 0.4$) and a visibility of $0.91$ for the first 21 LG modes; we refer the reader to Ref.\,\cite{Sevilla:2024_SpectralLG} for further details. After fiber coupling, one photon is sent to a monochromator (Andor Kymera 193i, FWHM $\approx 30$\,pm), and both photons are detected with single-photon avalanche diodes (SPADs). Single and coincidence counts are recorded using a time tagger. No correction was applied to the data in the measurements reported below.


In the first experiment, we set the pump's focusing waist to \( w_\mathrm{p} \approx 20 \, \mu\text{m} \) and projected the signal and idler into Gaussian modes with three different waist parameters: \( w_\mathrm{s}=w_\mathrm{i}\approx 10, 30, \) and \( 50 \, \mu\text{m} \). The spectrum for each configuration was recorded, as shown in the upper-right plot in Fig.\,\ref{fig:experiment_results}, where the starred points represent the experimental data and the solid line indicates the theoretical prediction. Notably, the measurements align with the previously mentioned effects associated with tighter focusing, including the central wavelength shift and asymmetry.

For the subsequent test, we fixed the focusing waist of the photons at \( w_\mathrm{s} = 30 \, \mu\text{m} \) and recorded the spectra of the joint modes \( P_{0,p_\mathrm{p}} \) for \( p_\mathrm{s} = 0, 1, 2 \). This was done under two different focusing conditions for the pump: \( w_\mathrm{p}= 60 \, \mu\text{m} \) (bottom-left plot) and \( w_\mathrm{p} = 20 \, \mu\text{m} \) (bottom-right plot). Once again, the results were accurately predicted by the developed theory.

\begin{figure}
    \centering
    \begin{overpic}[width=1.05\linewidth]{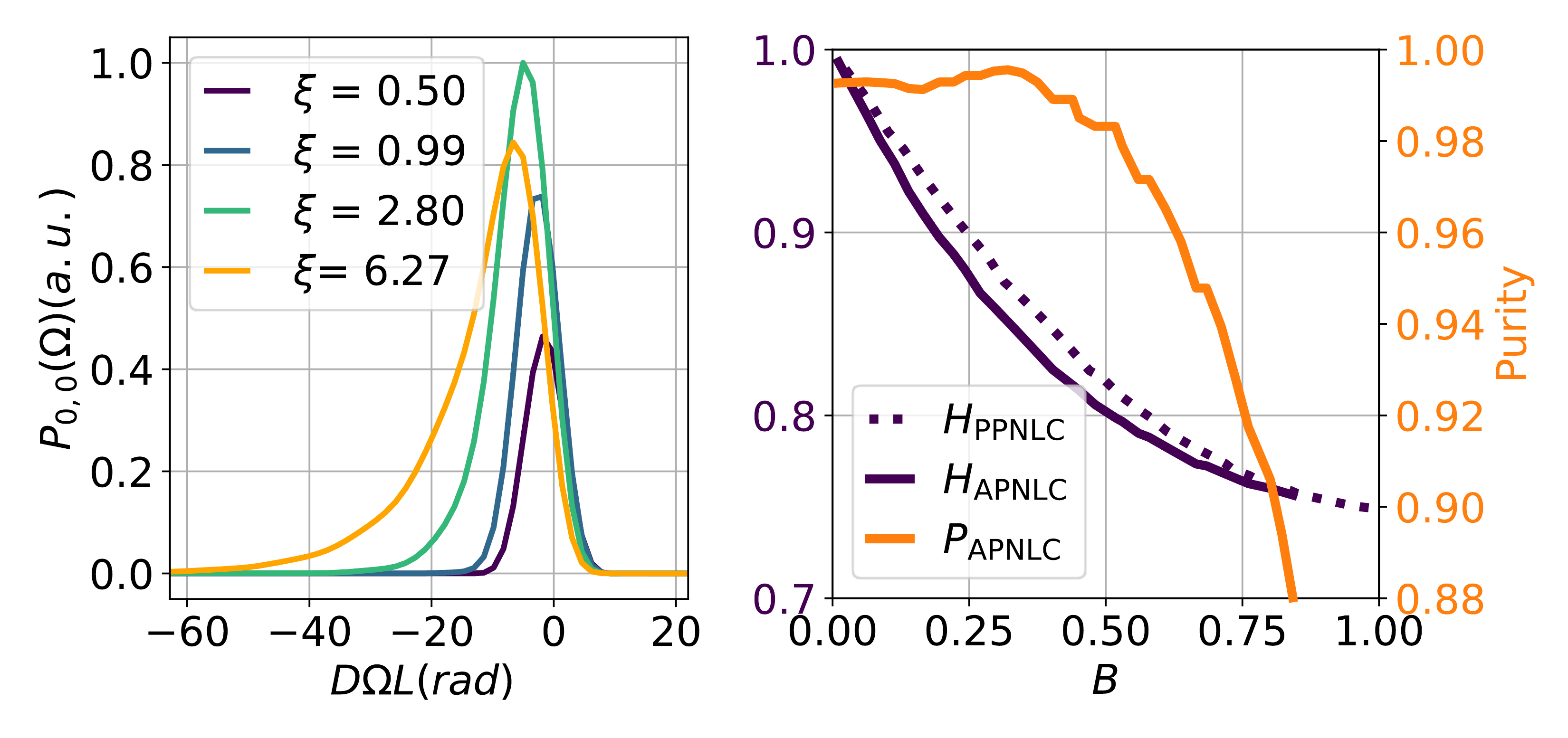}

   \end{overpic}
    
    \caption{(Left) Effect of focusing in the spectral profile $P_{0,0}(\Omega)$, showing spectral broadening and shift for tighter focusing. (Right) The solid and dashed blue lines, show the trade-off between heralding efficiency, $H_{max}$, and relative brightness for aperiodically poled (APNLC), and periodically poled (PPNLC) nonlinear crystals, respectively. The orange line, depicts the maximum spectral purity considering a transformed limited Gaussian beam. }
    \label{fig:apktp}
\end{figure}

\subsection{Type-II phase-matching with aperiodic poling}
Here we consider a Gaussian nonlinearity written as:

\begin{equation}\label{eq:gauss_NL}
    \chi^{(2)}(z)=\exp(-\frac{z^2}{2\sigma^2})\,.
\end{equation}

which can be performed experimentally via aperiodic poling of the nonlinear crystal \cite{Graffitti2017}. In the following, we refer to this configuration as an aperiodically poled nonlinear crystal (APNLC), in contrast to the periodically poled case (PPNLC). 

The primary motivation for employing a Gaussian nonlinearity is the enhanced spectral decorrelation of signal and idler photons, quantified by the spectral purity $\mathcal{P}$. We compute $\mathcal{P}$ assuming a spectrally Gaussian pump envelope $G(\Omega_\mathrm{s}+\Omega_\mathrm{i})=\exp{(-\frac{(\Omega_\mathrm{s}+\Omega_\mathrm{i})^2}{\sigma_\mathrm{p}^2}})$, where $\sigma_\mathrm{p}$ is the pump spectral bandwidth. In all of the following calculations $\sigma_\mathrm{p}$ is chosen so that $\mathcal{P}$ is maximized.  

Because the effective interaction length is reduced, we redefine the focusing parameter as

\begin{equation}
\xi=\frac{\sigma\lambda}{\sqrt{2\pi}n w^2},
\end{equation}

where $\sqrt{2\pi}\sigma$ corresponds to the effective crystal length obtained from the area under $\chi^{(2)}(z)$, which yields $\sqrt{2\pi}\sigma$ rather than $L$ for the case of $\chi^{(2)}=1$. We choose $\sigma=L/4$, which yields $\mathcal{P}>0.99$ for very loose focusing ($\xi_\mathrm{s}=\xi_\mathrm{p}=0.006$), approximating the plane-wave limit.

We then analyze the trade-off between relative brightness $B$ and heralding efficiency $H$ by optimizing $H$ over $\xi_\mathrm{p}$ and $\xi_\mathrm{s}$ at fixed $B$ (the third optimization strategy introduced above). For a direct comparison of the two types of poling, $B$ is normalized to the maximum pair-collection probability of the PPNLC case. The resulting trade-offs are shown in Fig.~\ref{fig:apktp}\,(right), where solid and dashed blue curves correspond to APNLC and PPNLC, respectively. While both configurations achieve the same maximal brightness, the APNLC exhibits a slightly less favorable trade-off at intermediate focusing, whereas the performance coincides in the limits of loose and tight focusing. We calculated the spectral similarity, following the same procedure as in Fig.\,\ref{fig:spectrum_Cp_Overlap}, using the focusing parameters obtained from the optimization of $H$ shown in Fig.\,\ref{fig:apktp}. We find that the APNLC case exhibits higher spectral distinguishability than the PPNLC case supporting the claim that the trade-off is originated due to the the spatial-spectral coupling.

Figure~\ref{fig:apktp}a illustrates the spectral broadening and asymmetry induced by tighter focusing. As in the periodically poled case, the maximum peak spectral density occurs near $\xi_\mathrm{s}=\xi_\mathrm{p}\approx2.84$, confirming the consistency of the rescaled focusing parameter. Deviations from a Gaussian spectral profile lead to a reduction in spectral purity, shown by the orange curve in Fig.~\ref{fig:apktp}\,(left). The purity drops rapidly for $B\gtrsim0.42$, corresponding to $\xi_\mathrm{s}\gtrsim1.4$.

\subsection{Collinear Type-0 periodic poling}\label{subsec:type-0}
Collinear type-0 SPDC differs qualitatively from type-II in that the longitudinal phase mismatch is dominated by the quadratic dispersion term, as the group-velocity mismatch vanishes ($D=0$). As a result, the residual phase mismatch $\phi$ plays a central role in shaping both the spectral and spatial properties of the emission and must be explicitly included in the evaluation of the relative brightness $B$ and heralding efficiency $H$.

\begin{figure}[t]
    \centering
    \includegraphics[width=1\linewidth]{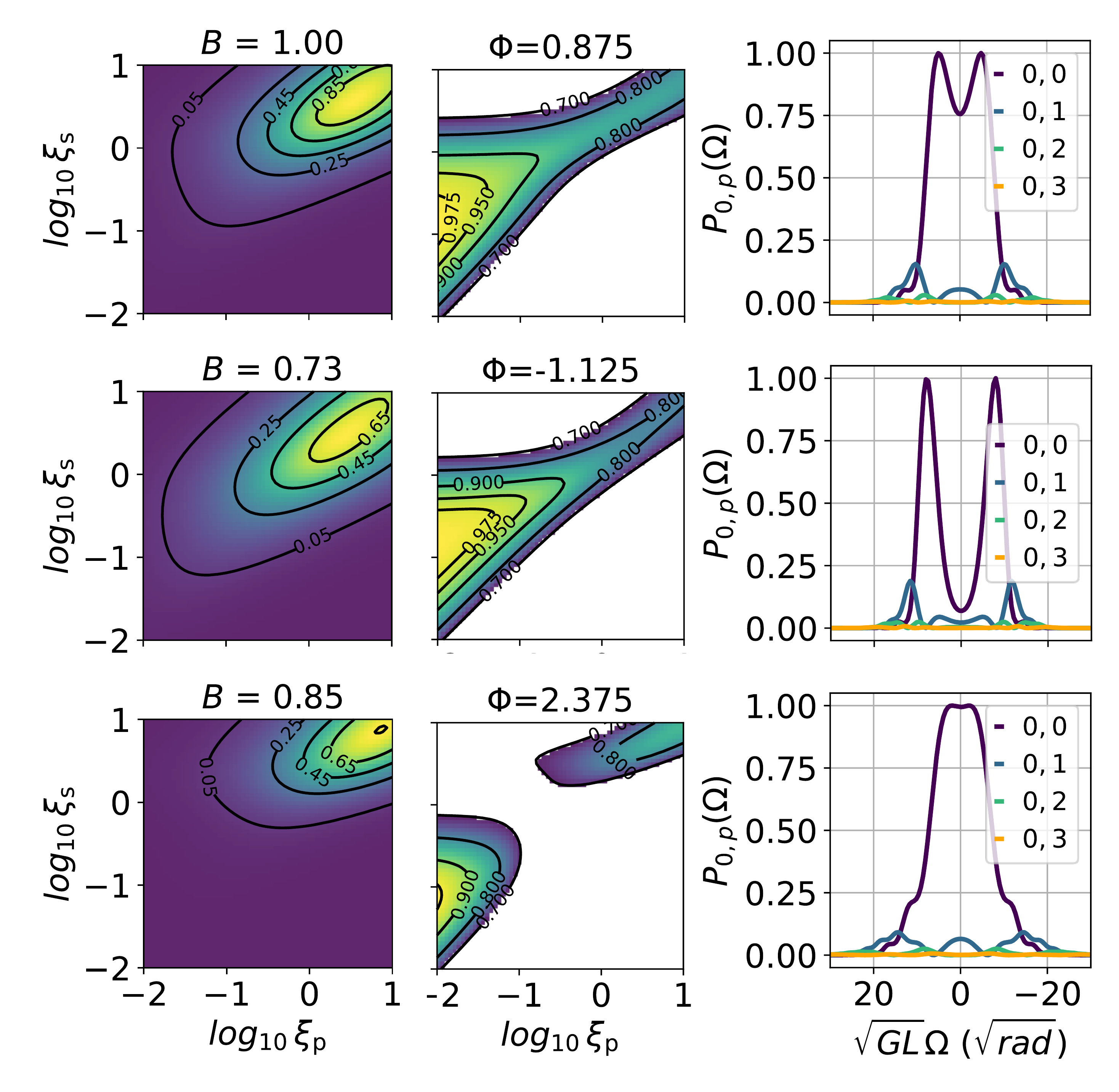}
    \caption{(Left column) Relative brightness and (center column) heralding efficiency as a function of $\xi_\mathrm{p}$ and $\xi_\mathrm{s}$, for three different values of phase mismatch ($\phi$). (Right column) Spectra of the joint spatial mode $P_{0,p}(\Omega)$ for $p=0,1,2,3$ for each case.}
        \label{fig:B_H_Spec_T0}
\end{figure}

Figure~\ref{fig:B_H_Spec_T0} shows $B$ and $H$ as functions of the pump and collection focusing parameters for three representative values of $\phi$. The case $\phi=0.875$ corresponds to the global maximum of $B$, achieved at $\xi_\mathrm{p}=3.98$ and $\xi_\mathrm{s,i}=3.55$. While a brightness-–heralding trade-off similar to that of type-II SPDC is observed, two key differences emerge: first, $B$ exhibits a well-defined maximum at finite focusing; second, a heralding efficiency of $H\approx0.85$ is attainable already at the brightest operating point, yielding a more favorable trade-off than in the type-II configuration.

\begin{figure}
    \centering
    \includegraphics[width=1\linewidth]{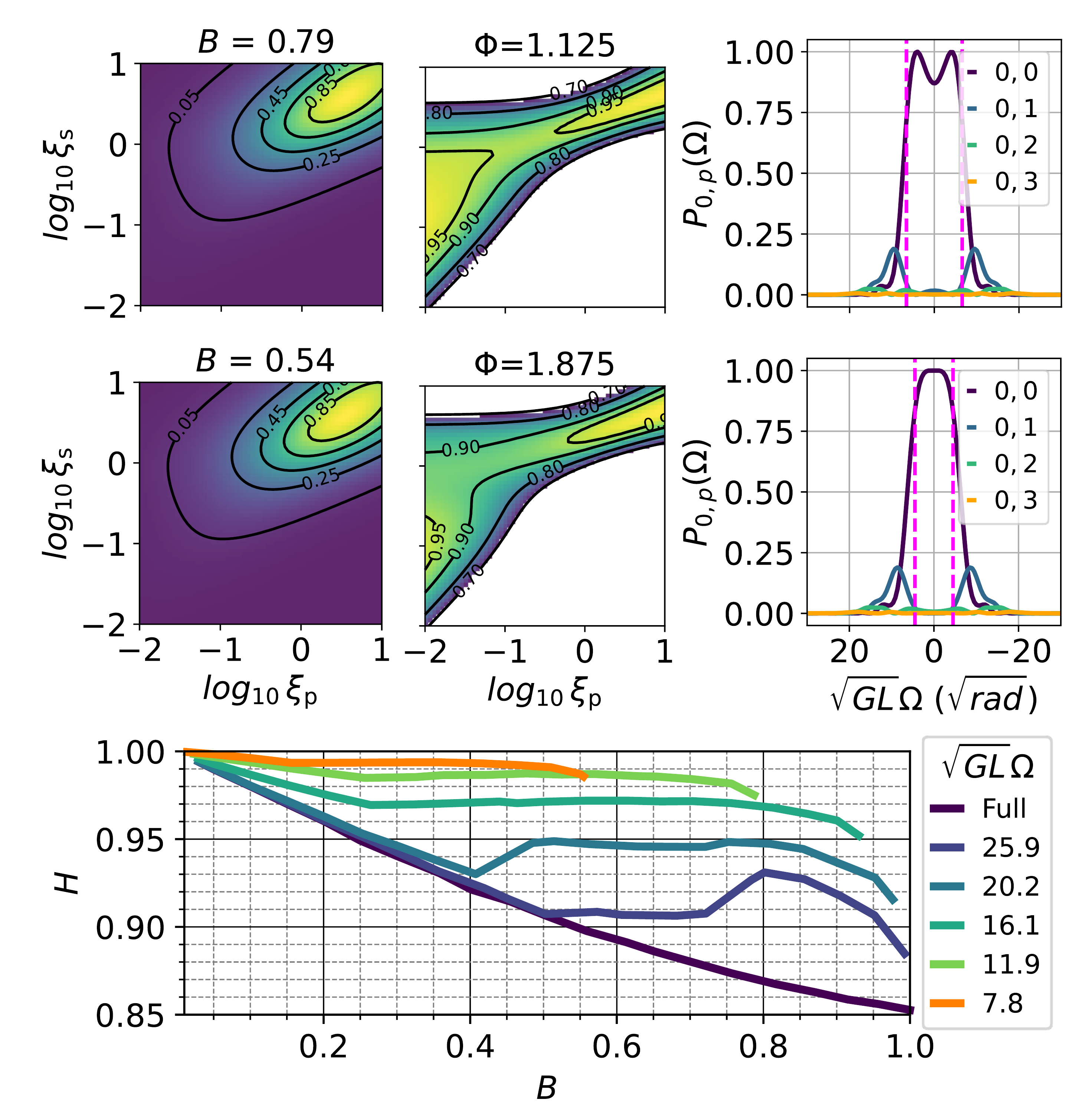}
    \caption{(Left column) Relative brightness and (center column) heralding efficiency as a function of $\xi_\mathrm{p}$ and $\xi_\mathrm{s}$, for two values of phase mismatch ($\phi$) when a gentle spectral filtering is applied. (Right column) $P_{0,p}(\Omega)$ for $p=0,1,2,3$ for each case, denoting the filtered region with dashed magenta lines. (Bottom) Optimization of $H$ over $\xi_\mathrm{s}$, $\xi_\mathrm{p}$ and $\phi$ for different values of $B$ and filter bandwidths.}
    \label{fig:B_H_Spec_T0_Spectral_filter}
\end{figure}

Varying the phase mismatch $\phi$ modifies both the maximum relative brightness (noted at the top of each subfigure of Fig.~\ref{fig:B_H_Spec_T0}) and the overall distribution of $B$ and $H$, especially noticeable in the lower row. In particular, the high-brightness region is shifted toward larger focusing parameters. This behavior can be understood from Eqs.\,\ref{eq:Long_Ph_mis_CW}: increased transverse momentum spread associated with tighter focusing partially compensates the phase mismatch.

The spectral origin of this behavior is illustrated by the joint-mode spectra $P_{0,p_\mathrm{s}}(\Omega)$ shown in the right column of Fig.~\ref{fig:B_H_Spec_T0}. For focusing parameters that maximize $B$, the fundamental mode dominates the central spectral region, while higher-order spatial modes primarily populate the spectral wings. This structure, combined with the intrinsically large bandwidth of type-0 SPDC, suggests that moderate spectral filtering can significantly enhance heralding efficiency with limited impact on brightness.

This is confirmed in Fig.~\ref{fig:B_H_Spec_T0_Spectral_filter}, which shows the effect of applying spectral filters that suppress higher-order mode contributions. For $\phi=1.125$, a filter bandwidth of $\sqrt{GL}\Delta\Omega' = 11.9\,\mathrm{rad}^{1/2}$ yields $H>0.98$ while maintaining $B>0.77$ at $\xi_\mathrm{p}=3.55$ and $\xi_\mathrm{s}=3.98$. To better visualize the effect of spectral filtering we plotted the spectral densities of the different higher-order joint modes to the right-hand side, where the dashed magenta lines depict the region of the spectral filtering. 

Another popular operation regime is shown for $\phi=1.875$, corresponds to a flat-top spectrum of the fundamental mode  (lower row in Fig.\,\ref{fig:B_H_Spec_T0_Spectral_filter}), that is particularly  relevant for frequency-domain entanglement\,\cite{CabrejoPonce2023}. In this configuration, the spectra are slightly more compressed around the degenerate wavelength, and the bandwidth of the fundamental mode decreases, resulting in a required narrower filtering.  Here, a narrower filter ($\sqrt{GL}\Delta\Omega' = 7.8\,\mathrm{rad}^{1/2}$) yields $H>0.98$ at $B=0.54$.

Finally, we summarize the trade-off in a similar way as for the type-II case. We optimize $H$ over $\xi_\mathrm{s}$, $\xi_\mathrm{p}$ and $\phi$ for different values of $B$ and filter bandwidths. The results are shown in Fig.\,\ref{fig:B_H_Spec_T0_Spectral_filter}(bottom). Here, the benefit of  spectral filtering is clear. Moderate spectral filtering can drastically improve the heralding efficiency without sacrificing the brightness significantly. For example,  curve $\sqrt{GL}\Delta\Omega' = 16.1\,\mathrm{rad}^{1/2}$ shows that it is possible to obtain $H\approx0.96$ and simultaneously maintain $B>0.9$, effectively eliminating any trade-off.

Although excessive spectral filtering is generally undesirable, it is commonly employed in type-0 sources due to their large bandwidth, which can otherwise degrade state purity in polarization-based schemes or in the presence of dispersive optical components. Moreover, frequency demultiplexing is a promising resource for multipartite quantum networks \cite{Brambila:23_ultrabiright}. Importantly, our results show that tight filtering is not required to achieve a measurable benefit: even moderate filtering suffices to achieve near-optimal heralding efficiency at or near maximum brightness. Note that additional considerations regarding the heralding efficiency apply in the pulsed-pump regime \cite{Meyer-Scott:2017_helrading_pulsed}.

\section{Conclusion}\label{sec:discussion}
In conclusion, our results highlight the frequency-dependent spatial mode structure as the underlying origin of the well-known trade-off between relative brightness and heralding efficiency in bulk SPDC, and translate this insight into quantitative and intuitive design rules for practical photon sources.

For collinear type-II SPDC, we find that optimizing the heralding efficiency rather than  brightness is generally the more favorable strategy. Furthermore, we find that at high focusing, the different joint spatial modes are spectrally distinguishable, fundamentally limiting the efficiency of coupling via spatial projection alone. We corroborated this experimentally using spatial- and spectral-projection measurements. We highlight that while narrowband spectral filtering can improve the trade-off, it is typically not experimentally viable since the  bandwidth of the source is often on the (sub-)nanometer scale. Extending the analysis to Gaussian nonlinearities realized via aperiodic poling, we show that the effect of the modified interaction can be captured by a simple rescaling of the focusing parameter. In this case, another experimentally crucial trade-off emerges: increasing the focusing of the SPDC photons reduces the spectral purity. We identify a threshold for the pair-collection and focusing parameter $B<0.42$ and $\xi_\mathrm{s}<1.4$ for which the spectral purity remains above $0.99$.

Applying the same formalism to type-0 SPDC reveals a qualitatively different fiber coupling behavior. By inspecting the spectra of higher-order joint spatial modes, appropriate spectral filtering bandwidth can be identified. We show how the trade-off evolves with different levels of spectral filtering and demonstrate that it can be almost completely mitigated with moderate filtering. For example, heralding efficiencies exceeding $H>0.98$ can be obtained without substantially compromising the brightness ($B>0.77$).

The present results apply to collinear, paraxial SPDC driven by single-mode Gaussian pump and collection fields in bulk nonlinear crystals. Extensions to non-paraxial regimes, spatial walk-off, and multi-crystal or interferometric source architectures constitute promising directions for future work.

\section*{Acknowledgements}
The authors thank Rene Sondenheimer for valuable discussions.  

\section{Funding}
The authors acknowledge support by the Carl-Zeiss-Stiftung within the Carl-Zeiss-Stiftung Center for Quantum Photonics (CZS QPhoton) under the project ID P2021-00-019. This work has received funding from the German Federal Ministry of Research, Technology and Space (BMFTR) within the PhoQuant project. We acknowledge support from the European Union’s Horizon 2020 Research and Innovation Action under Grant Agreement No.~899824 (SURQUID). CS and FS are members of the Max Planck School of Photonics, supported by the German Federal Ministry of Education and Research, the Max Planck Society, and the Fraunhofer Society. 

\section*{Data Availability}
The data and code are available in a deposit in Ref.\,\cite{zenodo_ppnlc_data}; (reserved DOI: 10.5281/zenodo.18851950); during peer review the deposit is accessible via a private link and will be published publicly upon acceptance/publication.

\appendix

\section{Theory}

The biphoton state of SPDC can be described by the expression:
\begin{align}\label{SPDC:freespace}
    \ket{\Psi} = \iint & d\bm{q}_\mathrm{s} \: d\bm{q}_\mathrm{i} \:d\omega_\mathrm{s} \: d\omega_\mathrm{i}\: \Phi(\bm{q}_\mathrm{s},\bm{q}_\mathrm{i},\omega_\mathrm{s},\omega_\mathrm{i})\nonumber\\&
  \hat{a}^{\dagger}_\mathrm{s}(\bm{q}_\mathrm{s},\omega_\mathrm{s})\:\hat{a}^{\dagger}_\mathrm{i}(\bm{q}_\mathrm{i},\omega_\mathrm{i})\ket{vac}.
\end{align}
where $\Phi({\bf q}_\mathrm{s},{\bf q}_\mathrm{i},\omega_\mathrm{s}, \omega_\mathrm{i})$ is the so-called two-photon amplitude (TPA). It encloses all the spatial-temporal correlations and is given by:

\begin{equation} \label{eq:modefunction}
\Phi({\bf q}_\mathrm{s},{\bf q}_\mathrm{i},\omega_\mathrm{s}, \omega_\mathrm{i})\sim  \int_{-L/2}^{L/2}dz\,\chi^{(2)}\,E_\mathrm{p}({\bf q_\mathrm{i}+q_\mathrm{s}},\omega_\mathrm{p}) e^{i\Delta kz}
\end{equation}
Here, $E_\mathrm{p}(\bf q_\mathrm{s}+q_\mathrm{i},\omega_\mathrm{p})$ is the amplitude of the pump beam in the spectral ($\omega_\mathrm{p}$) and transverse momentum ($\bf q_\mathrm{p}=q_\mathrm{s}+q_\mathrm{i}$) space, and the second term is the phase matching function (PMF), where $\Delta k= k^z_\mathrm{p}-k^z_\mathrm{s}-k^z_\mathrm{i}-2\pi/\Lambda$ is the phase mismatch between the longitudinal $k$-vectors of the interacting waves inside the nonlinear crystal with the poling period $\Lambda$. 

We can describe the spatial domain of Eq. ~\ref{SPDC:freespace} in the Laguerre-Gauss (LG) basis , which runs over the radial $p$ and azimuthal $\ell$ indices of the pump, signal and idler photons, leaving us with
\begin{align}\label{eq:decomposition}
    \ket{\Psi}=
    \sum_{p_\mathrm{s},p_\mathrm{i}=0}^{\infty} \sum^{\infty}_{\ell_\mathrm{p},\ell_\mathrm{s},\ell_\mathrm{i}=-\infty}\int d\Omega_\mathrm{s} d\Omega_\mathrm{i}\nonumber\\ 
    C_{p_\mathrm{p},p_\mathrm{s},p_\mathrm{i}}^{\ell_\mathrm{p},\ell_\mathrm{s},\ell_\mathrm{i}}(\Omega_\mathrm{s},\Omega_\mathrm{i}) \ket{p_\mathrm{s},\ell_\mathrm{s},\Omega_\mathrm{s}}\ket{p_\mathrm{i},\ell_\mathrm{i},\Omega_\mathrm{i}}.
\end{align}
Here, we have used the notation $\omega_j=\omega^0_{j}+\Omega_j$, where $\Omega_{j}$ is the frequency shift from the central frequency $\omega^0_{j}$. The probability amplitude $C_{p_\mathrm{p},p_\mathrm{s},p_\mathrm{i}}^{\ell_\mathrm{p},\ell_\mathrm{s},\ell_\mathrm{i}}(\Omega_\mathrm{s},\Omega_\mathrm{i})$ is the solution of the overlap integral

\begin{align}\label{coe1_appendix}
    C^{\ell_\mathrm{p},\ell_\mathrm{s},\ell_\mathrm{i}}_{p_\mathrm{p},p_\mathrm{s},p_\mathrm{i}}(\Omega_\mathrm{s},\Omega_\mathrm{i})
    =    \iint  d\bm{q}_\mathrm{s} \: d\bm{q}_\mathrm{i}
    \sinc(\frac{\Delta k(\bm{q}_\mathrm{s},\bm{q}_\mathrm{i},\Omega) L}{2})\nonumber&\\
    \times \mathrm{LG}_{p_\mathrm{p}}^{\ell_\mathrm{p}}(\bm{q}_\mathrm{p})[\mathrm{LG}_{p_\mathrm{s}}^{\ell_\mathrm{s}}(\bm{q}_\mathrm{s})]^*         [\mathrm{LG}_{p_\mathrm{i}}^{\ell_\mathrm{i}}(\bm{q}_\mathrm{i})]^*,
\end{align}

with $\mathrm{LG}_{p}^{\ell}(\bm{q}_\mathrm{s})$ defined as\,\cite{Baghi:2022_Generalized}:

\begin{align}\label{eq_sup:LG_function}
     \mathrm{LG}_{p}^{\ell}(\rho,\phi)=\mathrm{exp}\big(\frac{-\rho^2w^2}{4} +i\ell\phi\big)\sum_{u=0}^pT_u^{p,\ell}\rho^{2u+|\ell|},
\end{align}
with
\begin{equation}
         T_u^{p,\ell} = \sqrt{\frac{p!\,(p+|\ell|)!}{\pi}}\,
   \biggr(\frac{ w}{\sqrt{2}}\biggl)^{2u+|\ell|+1}\,\frac{(-1)^{p+u}(i)^{\ell}}{(p-u)!\,(|\ell|+u)!\,u!}
\end{equation}

Eq.\,\ref{coe1_appendix} is solved under the approximation:

\begin{equation}
k^z_\mathrm{j}=\sqrt{k^2_\mathrm{j}-|\bm{q_\mathrm{j}}|^2}\approx k_{\mathrm{j},0}+\frac{\Omega_\mathrm{j}}{u_{\mathrm{j}}}+\frac{G_{\mathrm{j}}\Omega_\mathrm{j}^2}{2}-\frac{|\bm{q_\mathrm{j}}|^2}{2k_{\mathrm{j},0}},
\label{eq:K_approx}
\end{equation}

applicable when $|\bm{q_\mathrm{j}}|\ll k_j$ and $\Omega_\mathrm{j}\ll\omega_{\mathrm{j},0}$. Here, $u_\mathrm{j}$ and $G_\mathrm{j}$  are the group velocity and the group velocity dispersion (GVD) at $\omega_{\mathrm{j},0}$, respectively. 

\section{Methods}

In this section, we summarize all the parameters, ranges and step size used for the calculation. Note that while we fix some of the parameters such as material, wavelengths and crystal length, the results were displayed in terms of adimensional parameters, such as $\xi$, $DL\Omega$ and $\sqrt{GL}\Omega$ making them applicable to different configurations. 

\subsection{Type-II quasi-phase-matching}\label{Appen-II}

For the numerical calculations, we consider the specific configuration of PPKTP crystal of length $L = 40\,mm$, pump wavelength $\lambda_\mathrm{p}=405\,nm$ and signal and idler wavelengths at $\lambda_\mathrm{s} = 810\,nm$. The range and step size of the parameters used in the calculations are given in the following table:

\begin{table}[h]
    \centering
    \begin{tabular}{|c|c|c|}
        \hline
        Parameter & Range & Step \\
        \hline
        $log_{10}(\xi)$ & -2 to 1 & 0.05 \\
        \hline
        $D\Omega L$ & -406.12 to 59.14 & 1.6 \\
        \hline
    \end{tabular}
    \label{tab:sampling_T2}
\end{table}

From the numerical data, we fit easy-to-use expressions for the optimization of the heralding as a function of the relative brightness:
\begin{equation}
    f(B)=\sum_{n=0}\alpha_nB^n
\end{equation}
where coefficients $\alpha_n$ are given in the following table: 
\begin{table}[h]
    \centering
    \begin{tabular}{|c|c|c|c|c|}
        \hline
        $f(B)$ & $\alpha_0$ & $\alpha_1$ & $\alpha_2$ & $\alpha_3$ \\
        \hline
        $H$ &  1.0008 & -0.4544 & 0.1561 & 0.0479 \\
        \hline
        $\log_{10}(\xi_\mathrm{p})$ & -1.9853 & 6.9212 & -9.8439 & 5.8524 \\
        \hline
        $\log_{10}(\xi_\mathrm{s})$ & -0.8787 & 3.7842 & -5.6981 & 3.7458\\
        \hline
    \end{tabular}
    \label{tab:fit_ppnlc}
\end{table}

\subsection{Type-II Gaussian phase matching}\label{app:Type-II_Gaussian}

The same as for PPNLC, we consider a crystal of length $L = 40\,mm$, pump wavelength $\lambda_\mathrm{p}=405\,nm$ and signal and idler wavelengths at $\lambda_\mathrm{s} = 810\,nm$. The range and steps size of the parameters used in the calculations are given in the following table:

\begin{table}[h]
    \centering
    \begin{tabular}{|c|c|c|}
        \hline
        Parameter & Range & Step \\
        \hline
        $log_{10}(\xi)$ & -2.2 to 0.8 & 0.05 \\
        \hline
        $D\Omega L$ & -406.12 to 59.14 & 1.6 \\
        \hline
    \end{tabular}
    \label{tab:sampling_T2_APNLC}
\end{table}

From the numerical data, we fit easy-to-use expressions for the optimization of the heralding as a function of the relative brightness:
\begin{equation}
    f(B)=\sum_{n=0}\alpha_nB^n
\end{equation}
where coefficients $\alpha_n$ are given in the following table: 
\begin{table}[h]
    \centering
    \begin{tabular}{|c|c|c|c|c|}
        \hline
        $f(B)$ & $\alpha_0$ & $\alpha_1$ & $\alpha_2$ & $\alpha_3$ \\
        \hline
        $H$ &  1.0052 & -0.6337 & 0.5418 & -0.1658\\
        \hline
        $\log_{10}(\xi_\mathrm{p})$ &  -2.0472 & 8.921 & -14.4135  & 9.1454 \\
        \hline
        $\log_{10}(\xi_\mathrm{s})$ & -0.8802 & 4.6923 & -7.3127 & 4.9145\\
        \hline
    \end{tabular}
    \label{tab:fit_apnlc}
\end{table}

\subsection{Type-0 quasi-phase-matching}

For the numerical calculations, we consider the specific configuration of PPKTP crystal of length $L = 5\,mm$, but keep the same parameters for $\lambda_\mathrm{p}=405\,nm$ and $\lambda_\mathrm{s} = 810\,nm$. The range and step size of the parameters used in the calculations are given in the following table:

\begin{table}[h]
    \centering
    \begin{tabular}{|c|c|c|}
        \hline
        Parameter & Range & Step \\
        \hline
        $log_{10}(\xi_{p,s,i})$ & -2 to 1 & 0.05 \\
        \hline
        $\sqrt{GL}\Omega $ & -41.157 to 35.065 & 0.518 \\
        \hline
        $\phi $ & -1.125 to 2.375 & 0.25 \\
        \hline
    \end{tabular}
    \label{tab:sampling_T0}
\end{table}

\section{ Extended discussion}
\subsection{Scaling of relative brightness with crystal length}\label{app_scaling}

The relative brightness $B$ is directly proportional to the probability of generating both photons in the Gaussian mode proportional to $\int d \Omega P_{0,0}(\Omega)$. Here $P_{0,0}(\Omega)$ is the magnitude squared of the spectral amplitude $P_{0,0}(\Omega) = |C_{0,0}(\Omega)|^2$ and is defined as:

\begin{align}\label{app_eq:diff0}
    P_{0,0}(\Omega) = 
    \Big|L\int d\rho_\mathrm{s}d\rho_\mathrm{i}d\phi_\mathrm{s}d\phi_\mathrm{i}\mathcal{F}(\rho_\mathrm{s},\rho_\mathrm{i},\phi_\mathrm{s},\phi_\mathrm{i},\Omega,L)\Big|^2\,.
\end{align}

which was written in the polar-coordinate representation of the transverse momentum. The function $\mathcal{F}$ inside the integral is given by\,\cite{Baghi:2022_Generalized}:

\begin{align}\label{app_eq:diff1}
    \mathcal{F} 
    = \rho_\mathrm{s}\rho_\mathrm{i}\exp\Big[-\frac{1}{4}\Big(\rho_\mathrm{p}^2w_\mathrm{p}^2+\rho_\mathrm{s}^2w_\mathrm{s}^2+\rho_\mathrm{i}^2w_\mathrm{i}^2\Big)\Big]   \nonumber&\\
    \sinc\Big[\Big(\frac{k_\mathrm{p}-k_\mathrm{s}}{2k_\mathrm{p}k_\mathrm{s}}\rho_\mathrm{s}^2+\frac{k_\mathrm{p}-k_\mathrm{i}}{2k_\mathrm{p}k_\mathrm{i}}\rho_\mathrm{i}^2 \nonumber&\\- \frac{\cos(\phi_\mathrm{s}-\phi_\mathrm{i})}{k_\mathrm{p}} \rho_\mathrm{s}\rho_\mathrm{i}   -O(\Omega)\Big)\frac{L}{2}\Big] ,
\end{align}

All spectral dependencies have been grouped into $O(\Omega)= D\Omega+G\Omega^2$, as defined in the main text. Applying the following change of variables $\rho_j=\sqrt{L}\,\rho_j'$ and $w_j^2 = \frac{L}{k_j \xi_j^2}$ (note that $\xi_j$ is the focusing parameter), and noticing that $\rho_\mathrm{p}^2 = \rho_\mathrm{s}^2 + \rho_\mathrm{i}^2 +2\rho_\mathrm{s}\rho_\mathrm{i}\cos(\phi_\mathrm{s}-\phi_\mathrm{i})$,  Eqs.\,\ref{app_eq:diff0} and \ref{app_eq:diff1}. can be re-written as:

\begin{align}\label{app_eq:diff2}
    P_{0,0}(\Omega) = 
    \Big|L\Big(\frac{1}{\sqrt{L}}\Big)\int d\rho_\mathrm{s}'d\rho_\mathrm{i}'d\phi_\mathrm{s}d\phi_\mathrm{i}\mathcal{F}(\rho_\mathrm{s}',\rho_\mathrm{i}',\phi_\mathrm{s},\phi_\mathrm{i},\Omega,L)\Big|^2\,.
\end{align}

and, 

\begin{align}\label{app_eq:diff3}
    \mathcal{F} 
    = \rho_\mathrm{s}'\rho_\mathrm{i}'\exp\Big[\frac{k_\mathrm{p}\xi_\mathrm{p}+k_\mathrm{s}\xi_\mathrm{s}}{4k_\mathrm{p}k_\mathrm{s}\xi_\mathrm{p}\xi_\mathrm{s}}\rho_\mathrm{s}'^2+\frac{k_\mathrm{p}\xi_\mathrm{p}+k_\mathrm{i}\xi_\mathrm{i}}{4k_\mathrm{p}k_\mathrm{i}\xi_\mathrm{p}\xi_\mathrm{i}}\rho_\mathrm{i}'^2 \nonumber&\\    
    + \frac{\cos(\phi_\mathrm{s}-\phi_\mathrm{i})}{2k_\mathrm{p}\xi_\mathrm{p}} \rho_\mathrm{s}'\rho_\mathrm{i}'  \Big]  \nonumber&\\    
    \sinc\Big[\frac{k_\mathrm{p}-k_\mathrm{s}}{4k_\mathrm{p}k_\mathrm{s}}\rho_\mathrm{s}'^2+\frac{k_\mathrm{p}-k_\mathrm{i}}{4k_\mathrm{p}k_\mathrm{i}}\rho_\mathrm{i}'^2 \nonumber&\\    
    - \frac{\cos(\phi_\mathrm{s}-\phi_\mathrm{i})}{2k_\mathrm{p}} \rho_\mathrm{s}'\rho_\mathrm{i}'   -O(\Omega)\frac{L}{2}\Big] ,
\end{align}

respectively. From this we can see that diffraction adds and extra dependency on the crystal length, namely ($L^{-1}$), when considering the same focusing parameters. Note that keeping $\xi_j$ fixed, implies adjusting the waists parameters.  

We now integrate $P_{0,0}(\Omega)$ over $\Omega$, introducing a change of variables adapted to the corresponding spectral dependence. This is done separately for type-II $(D \gg G)$ and degenerate type-0 $(D = 0)$ SPDC, since $O(\Omega) \approx D\Omega$ in the former case, whereas $O(\Omega) \approx G\Omega^2$ in the latter. Accordingly, we define $\Omega = \Omega'/L$ and $\Omega = \Omega'/\sqrt{L}$, respectively, canceling the $L$ dependency in the argument of the Sinc function.


It is easy to see that the scaling of the brightness with crystal length is given by: 
\begin{align}\label{app_eq:diff5}
    B_{D>>G}    \propto    L^2\Big(\frac{1}{L}\Big)_{spat} \Big(\frac{1}{L}\Big)_{spect} F &\\
    B_{D=0}    \propto    L^2\Big(\frac{1}{L}\Big)_{spat} \Big(\frac{1}{\sqrt{L}}\Big)_{spect} F 
\end{align}

where the spatial and spectral contribution to the scaling have been emphasized. All remaining $L$- independent terms have been grouped into $F$ which is defined as:

\begin{align}\label{app_eq:diff6}
    F = \int d\Omega'\Big|\int d\rho_\mathrm{s}'d\rho_\mathrm{i}'d\phi_\mathrm{s}d\phi_\mathrm{i}\mathcal{F}(\rho_\mathrm{s}',\rho_\mathrm{i}',\phi_\mathrm{s},\phi_\mathrm{i},\Omega')\Big|^2\,. 
\end{align}

In the following table, we show a summary of the scaling of the brightness with respect to the crystal length for waveguide and bulk crystals and for different phase matching configurations. Furthermore, we also add for the case of unfiltered and filtered cases. For the latter, the filter bandwidth is much lower than the SPDC bandwidth. Importantly, for the scaling in a bulk crystal to hold, it considers a fixed focusing parameter as can be seen from \ref{app_eq:diff3}. This means that the waists of the Gaussian modes need to be appropriately adjusted.  

\begin{table}[t]
    \centering
    \begin{tabular}{|c|c|c|}
        \hline
        Type & unfiltered & filtered  \\

        \hline
        Waveguide $D\gg G$& $L$ & $L^2$ \\
        \hline
        Waveguide $D=0$& $L^{3/2}$ & $L^2$ \\
        \hline 
        Bulk $D\gg G$& $const$ & $L$ \\
        \hline
        Bulk $D=0$& $\sqrt{L}$ & $L$ \\
        \hline
    \end{tabular}
    \label{tab:scaling_L}
\end{table}

\subsection{Equal signal and idler focusing in degenerate and nondegenerate SPDC}
\label{sec:appen_nondeg}

The spatial correlations of SPDC photon pairs are determined by the product of the pump envelope and the phase-matching function:

\begin{align}\label{app_eq:focus1}
    \Phi
    = \exp\Big(-\frac{|\mathbf{q}_\mathrm{p}|^2w_\mathrm{p}^2}{4}\Big)   
    \sinc\Big[\Big(    \frac{|\bm{q_\mathrm{p}}|^2}{k_{\mathrm{p}}}  - \frac{|\bm{q_\mathrm{s}}|^2}{k_{\mathrm{s}}}  - \frac{|\bm{q_\mathrm{i}}|^2}{k_{\mathrm{i}}}                \Big)\frac{L}{4}\Big] ,
\end{align}

Introducing the focusing parameter, and considering the degenerate case
\(\lambda_\mathrm{s}=\lambda_\mathrm{i}=\lambda_\mathrm{p}/2\), together with the approximation
\(k_\mathrm{p}\approx 2k_\mathrm{s}\approx 2k_\mathrm{i}\), Eq.~\eqref{app_eq:focus1} can be written as~\cite{Palacios:11}

\begin{align}\label{app_eq:focus2}
    \Phi
    \approx \exp\Big(-\frac{L}{4}\frac{|\bm{q_\mathrm{p}}|^2}{k_{\mathrm{p}}} \xi_\mathrm{p}^{-1} \Big) 
    \sinc\Big(    \frac{|\bm{q_\mathrm{m}}|^2}{k_{\mathrm{p}}}                \frac{L}{4}\Big) \,.
\end{align}

Here, $\bm{q_\mathrm{m}}=\bm{q_\mathrm{s}}-\bm{q_\mathrm{i}}$ and $\bm{q_\mathrm{p}}=\bm{q_\mathrm{s}}+\bm{q_\mathrm{i}}$. Equation~\eqref{app_eq:focus2} shows that, when written in terms of the focusing parameter, the transverse momentum spread of the pump is rescaled by the factor $L/(4k_\mathrm{p})$. This makes the angular bandwidth independent of $L$ and $\lambda_\mathrm{p}$ in the normalized variables. Moreover, in this degenerate limit the phase-matching function depends only on $\bm{q}_\mathrm{m}$, whereas the pump envelope depends only on $\bm{q}_\mathrm{p}$. This symmetry suggests that the signal and idler photons have similar angular mode profiles, making the choice $\xi_\mathrm{s}=\xi_\mathrm{i}$ a natural and physically well-motivated choice.

For type-II or nondegenerate configurations, the spatial dependence of $\Delta k$ can be written as

\begin{equation}
\label{app_eq:focus3_nondeg}
\begin{split}
\Delta k
&\approx \frac{|k_\mathrm{s}\bm{q}_\mathrm{i}-k_\mathrm{i}\bm{q}_\mathrm{s}|^2}
{2k_\mathrm{p}k_\mathrm{s}k_\mathrm{i}} \\
&= \frac{|(k_\mathrm{s}+k_\mathrm{i})\bm{q}_\mathrm{m}+(k_\mathrm{s}-k_\mathrm{i})\bm{q}_\mathrm{p}|^2}
{8k_\mathrm{p}k_\mathrm{s}k_\mathrm{i}} \,.
\end{split}
\end{equation}

where the approximation $k_\mathrm{p}=k_\mathrm{s}+k_\mathrm{i}$ has been used.

In this case, the transverse momentum phase matching no longer acts solely along the $\bm{q}_\mathrm{m}$ axis, because the signal and idler transverse momenta are weighted by their respective wave numbers $k_j$. For degenerate type-II SPDC, the birefringence is typically small and only weakly perturbs this symmetry. For strongly nondegenerate configurations, however, this rescaling must be treated more carefully; such configurations are therefore outside the scope of this work.

\bibliography{coupling-paper}

\end{document}